\documentclass{ws-ijgmmp}
\usepackage[super]{cite}
\usepackage{graphicx}
\begin{document}

\markboth{V. Bhardwaj, A. K. Yadav, L. K. Gupta, R. Prasad, S. Srivastava}
{Constraining hybrid potential scalar field.....}

%%%%%%%%%%%%%%%%%%%%% Publisher's Area please ignore %%%%%%%%%%%%%%
\catchline{}{}{}{}{}
%%%%%%%%%%%%%%%%%%%%%%%%%%%%%%%%%%%%%%%%%%%%%%%%%%%%%%%%%%%%%%%%%%%
%%%%%%%%%%%%%%%%%%%%%%%%%%%%%%%%%%%%%%%%%%%%%%%%%%%%%%%%%%%%%%%%%%%%

\title{Constraining hybrid potential scalar field cosmological model in Lyra's geometry with recent observational data}

\author{Vinod Kumar Bhardwaj}

\address{Department of Mathematics, Department of Mathematics, GLA University\\ Mathura-281 406, India\\
\email{dr.vinodbhardwaj@gmail.com}}

\author{Anil Kumar Yadav}

\address{Department of Physics, United College of Engineering and
Research\\ Greater Noida - 201310, India\\
\email{abanilyadav@yahoo.co.in}}

\author{Lalit Kumar Gupta}

\address{Department of Physics, Shyamlal saraswati Mahavidyalaya (PG), Shikarpur, Bulandshahr - 203395, India\\
\email{lalitl247@gmail.com}}

\author{Rajendra Prasad}

\address{Department of Physics, Galgotias College of Engineering and Technology, Greater Noida - 201310, India\\
\email{drrprasad@gmail.com}}

\author{Sudhir Kumar Srivastava}

\address{Department of Mathematics, Janata Janardan Inter College, Gandhi Nagar\\ Ghazipur 233225, India\\
\email{drsudhirgzp@gmail.com}}

\maketitle

%\pub{Received (Day Month Year)}{Revised (Day Month Year)}

\begin{abstract}
In the current study, we investigate a scalar field cosmological model with Lyra's geometry to explain the present cosmic expansion in a homogeneous and isotropic flat FRW universe. In Einstein's field equations, we presupposed a variable displacement vector as an element of Lyra's geometry. In the context of the conventional theory of gravity, we suggest a suitable parameterization of the scalar field's dark energy density in the hybrid function of redshift $z$, confirming the essential transition behavior of the universe from a decelerating era to the present accelerated scenario. We present constraints on model parameters using the most recent observational data sets from OHD, BAO/CMB, and Pantheon, taking Markov Chain Monte Carlo (MCMC) analysis into account. For the proposed model, the best estimated values of parameters for the combined dataset (OHD, BAO/CMB, and Pantheon) are $ H_0 = 71.15\pm 0.26$ km/s/Mpc, $ \Omega_{m0}=0.2625\pm 0.0024$, $ \Omega_{\phi0} = 0.676\pm0.038$, $ \alpha=-0.22\pm0.13$, $n = 0.096\pm0.079$, and $k = 0.38\pm0.32$. The model exhibits a flipping nature, and the redshift transition occurs at $z_t = 0.756^{+0.005}_{-0.015}$. The current value of the decelerated parameter for the proposed model is calculated as $q_0 = -0.625^{+0.067}_{-0.085}$ for the combined dataset. Some dynamical properties of the model like energy density ($\rho_{\phi}$), scalar field pressure ($p_{\phi}$), EoS parameter of scalar field ($\omega_{\phi}$), and effective EoS parameter ($\omega_{eff}$) are analyzed and presented. Further, we have also examined the statefinder diagnosis and jerk parameters of the derived model. The total density parameter for the derived model is found to be unity which is in nice agreement with recent standard findings.
\end{abstract}

\keywords{FRW space-time; Scalar field theory; Observational constraints; Lyra geometry.}
%%%%%%%%%%%%%%%%%%%%%%%%%%%%%%%%%%%%%%%%%%%
\section{Introduction}
In order to explain how gravity interacts with space and time, Einstein developed the general relativity (GR) theory at the beginning of the 20th century. He makes a clear connection between space-time geometry and matter and radiation. Relating the energy and momentum, fundamental characteristics of matter and radiation, and the curvature of space-time, Einstein presented the field equation for GR as $R_{ij}-\frac{1}{2} R g_{ij}=8 \pi G T_{ij}$ \cite{ref1}. Since the theoretical development of general relativity that connect the geometry with gravity, several approaches and theories have been proposed in search of additional geometrization models. 
The idea in which the geometrization of gravity with fundamental forces has also been proposed. In this direction, Weyl suggested the idea of geometrizing gravity and electromagnetic together using Riemannian geometry to unite them under the umbrella of a ``unified field theory" \cite{ref2}. Weyl suggested the usage of Riemannian geometry and vector potential to describe electromagnetic forces. In general, as a vector goes across space, the net displacement is determined by the initial and ending positions, ignoring the history of the path travelled results in the non-integrability of path followed. Weyl's approach of using vector potential complicated the issue and hence debarred \cite{ref2,ref3}. In the subsequence, several additional theories have also been proposed, either to replace Einstein's theory of relativity or to reduce the complexity of Weyl's theory. Lyra proposed the use of the Gauge function with Riemannian geometry in the progress \cite{ref4}. Lyra's idea resembles with Weyl's method and maintains the integrability of length transfer, as would be expected in Riemannian geometry. Since Lyra's geometrization behaves in accordance with Einstein's principle, it has frequently been used by many researchers to predict and explain cosmic phenomena \cite{ref5,ref6,ref7,ref8}.  In Lyra's context, Sen suggested a static cosmic model behaving like Einstein's model under static conditions, though model suffers red shift \cite{ref9}. On the other hand, Halford used Lyra's geometry to explore non-static cosmic events. Halford also pointed that a constant Gauge vector field developed when the Gauge function was introduced into Lyra's geometry and behaves similarly to the cosmological constant $\Lambda$\cite{ref10}. Imposing the Lyra’s suggestions, Soleng, additional explored the importance of Gauge vector field as a source of creation in cosmic theories \cite{ref8}. Under the observational limits, several cosmological theories are developed on the basis of Lyra's geometry depicting similar consequences as anticipated in Einstein's theory of relativity \cite{ref11,ref12,ref13}. Moreover, the authors of Refs. \cite{Yousaf/2024,Malik/2024,Malik/2024a,Bhatti/2023} have investigated some useful applications of modified theories of gravity in various physical contexts. In particular, Yousaf et al. \cite{Yousaf/2024} have analyzed an axially symmetric radiating system in f(R) gravity. Malik et al. \cite{Malik/2024,Malik/2024a} have investigated singularity-free embedding stellar structures and bouncing universe with cosmological parameters in modified theories of gravity respectively while in Bhati et al. \cite{Bhatti/2023}, the electromagnetic effect on compact object in $f(R,T)$ theory of gravity is described.\\

Several astrophysical experiments have confirmed that the cosmos is expanding at an accelerated rate in present time \cite{ref14,ref15,ref16,ref17}. According to many recent studies \cite{ref18,ref19,ref20}, dark energy (DE) is anticipated to play a major role in the universe's expansion, whereas dark matter is anticipated to be a key component in the growth of large-scale structures (LSS). The mysterious kind of energy called as DE, which exerts a tremendous amount of repulsive (negative) pressure, is what causes the cosmos to expand. With the experimentally confirmations of present cosmic reality, theoretical researchers are motivated to create universe models in various frameworks. The cosmological term has been believed to be a suitable replacement of DE because of its repulsive behavior \cite{ref21}. In literature, to explain the universe's present accelerated expansion, a variety of alternative hypotheses without the cosmological constant (CC) have been suggested. Each of these theories has a different prediction in describing the characteristics of DE and cosmic behaviour of the universe. In order to fit observational data, some of these theories also have extra parameters that can be adjusted. Although the cosmological constant matches the scientific results well and is validated by several experiments, it has failed to characterize the inflationary period of the cosmos. In addition to modified theories, several scalar field models are also introduced in theoretical cosmology, to address these issues of inflationary age and to describe the current expanding era of the cosmos \cite{ref22,ref23}.\\

In these studies, the scalar field ($\phi$) is considered as an assumption for the dark energy component which produces the negative pressure along with a reducing potential ($V(\phi)$). In literature, various cosmological research depending on the scalar field is suggested to characterize the dynamics of the cosmos \cite{ref22,ref23,ref24,ref25}. The quintessence is an interesting scalar field model that precisely avoids the conventional issues of fine-tuning and cosmic coincidence and depicts the present cosmic reality \cite{ref24,ref25}. Johri \cite{ref26} was the first to propose the idea of tracking, indicating a certain direction to explain the current cosmic scenario using the potential of the tracker. This idea was strongly supported by the observational estimates. Numerous quintessence models have been proposed in literary works. For a non-minimal relationship between dark matter and quintessence\cite{ref27,ref28,ref29}, these concepts include the potential for a scalar field to evolve under the influence of an unconventional kinetic term. The important applications of a variable EoS (Equation of State) parameter in the framework of scalar-tensor theory can be seen in Ref. \cite{ref30,ref31}. The existence of the scalar field in astrophysical investigations is also acknowledged by several fundamental theories. Numerous cosmic models have recently been developed in various scalar field theory frameworks \cite{ref30,ref31,ref32,ref33,ref34}. Kamenshchik et al. examined the Chaplygin gas DE model dark with the aid of a special form of EoS \cite{ref35}.\\

In the current study, we investigated a scalar field cosmological model with Lyra's geometry to explain the present cosmic expansion in a homogeneous and isotropic flat FRW universe. In Einstein's field equations, we assumed a variable displacement vector as an element of Lyra's geometry. The model parameters are extracted using the most recent observational data sets from OHD, BAO/CMB, and Pantheon. The manuscript is developed in the following manner. The model and its solutions taking hybrid scalar field density are described in section 2. Observational data and methodology for constraining the model parameters are mentioned in section 3. The features and dynamical characteristics of the model are discussed in section 4. A brief concluding summary of the proposed model is declared in section 5. 
%%%%%%%%%%%%%%%%%%%%%%%%%%%%%%%%%%%%%%%%%%%%%%%%%%%%%%%%%%%%%%%%
\section{Field equations and its solution }
Following Sen \cite{ref9}, we consider the action proposed for gravitational field equations in Lyra's geometry. 
\begin{equation}
S_{\psi} =  \int {d^4 x \sqrt{-g} \bigg[ \frac{1}{2} \psi^4 R +\mathcal{L}_{m} \bigg]}
\end{equation}
where $\mathcal{L}_{m}$ stands for matter Lagrangian and $8 \pi G = 1 = c$.\\ 
The field equations in Lyra's geometry \cite{ref3,ref4,ref5,ref9,ref10,ref35a} are recast as
\begin{equation}
R_{i j} -\frac{1}{2} R g_{i j}+\frac{3}{4}\psi_{i}\psi_{j}-\frac{3}{2}\psi_{i}\psi^{i}=T_{ij}
\end{equation}
where, perfect fluid's energy-momentum tensor $T_{i j}$ is described by $T_{ij} = \frac{-2}{\sqrt{-g}} \frac{\delta (\sqrt{-g} \mathcal{L}_{m})}{\delta g^{ij}}$, $R_{ij}$ represents the Ricci tensor, scalar curvature is denoted by R, and  displacement vector $\psi_{i}$ is the function of time and is defined as $\psi_{i}=(\beta(t),0,0,0)$.\\
%For the purpose of modeling universe's large-scale structure and to study its evolutionary geometry, we consider a 4-dimensional FRW space-time universe which is flat, homogeneous, and isotropic in nature.
For the purpose of modelling structure of cosmos and to study its evolutionary geometry, we consider a 4-dimensional FRW space-time universe which is flat, homogeneous, and isotropic in nature.
\begin{equation}
ds^2= a(t)^2 \left( dx^2+dy^2+ dz^2 \right)-dt^2
\end{equation}  
%where average scale factor $a(t)$ is used to estimate cosmic growth of universe with time. For the above line element, the Ricci scalar can determined in the form $R = - (\dot{H}+2H^2)$, here $H$ is the Hubble parameter defined as $H=\frac{\dot{a}}{a}$. In our study, we assumed that the universe has a flat geometry because this is a usual forecast of the inflationary model which is also confirmed by several experimental observations, such as the LSS surveys and the CMB measurements \cite{ref16,ref20,ref36,ref37}. 
%The flat universe model gets more attention for cosmological studies at present because it is simple and it only needs few extra parameters than the $\Lambda$CDM base model.\\
where average scale factor $a(t)$ is used to estimate the cosmic growth of the universe with time. For the above line element, the Ricci scalar can determined in the form $R = - (\dot{H}+2H^2)$, here $H=\frac{\dot{a}}{a}$ is the Hubble parameter. In our study, we assumed that the universe has a flat geometry because this is a usual forecast of the inflationary model which is also confirmed by several experimental observations, such as the LSS surveys and the CMB measurements \cite{ref16,ref20,ref36,ref37}.
The flat universe model gets more attention for cosmological studies at present because it is simple and it only needs a few extra parameters than the $\Lambda$CDM base model.\\
%Furthermore, since it only needs few extra parameters than the $\Lambda$CDM base model, a flat universe also has the advantage of being the most simple and plausible model.\\

%In our study, we assumed that the universe has a flat geometry because this is a usual prediction of the inflationary paradigm and is also confirmed by several cosmological observations, such as the LSS surveys and the CMB measurements [7,8, 5,6]. Furthermore, since it only needs few extra parameters than the $\Lambda$CDM base model, a flat universe has the advantage of being the most straightforward and reasonable model.\\
For an ideal fluid, the tensor of energy-momentum can be recasts in terms of energy density, velocity, and fluid pressure as $T^{m}_{ij} = (p_{m}+\rho_{m}) u_{i} u_{j}-p_{m} g_{ij}$, where $\rho_{m}$ and $p_{m}$ are the energy density and pressure of the matter.\\
In co-moving coordinate system, the field equations (2) for metric (3) are developed as
\begin{equation}
3 H^2 -\frac{3}{4} \beta^2 = \rho_{m}
\end{equation}
\begin{equation}
2\dot{H}+3 H^2 +\frac{3}{4} \beta^2 = -p_{m}
\end{equation}
%A mathematical statement that demonstrates the scalar field's interaction with gravity provides its action. A fictitious field called the scalar field has been proposed to explain a variety of physics phenomena, including inflation, DE, and Higgs technique. The action try to generalize the theory of GR using basic concepts of scalar-tensor gravitational theories.  In the situation, the scalar field is vital in adjusting the gravitational force's strength, which results in a wide range of physical events. Usually, for a scalar field the action is expressed in terms of scalar field and its derivatives as
A mathematical statement that demonstrates the scalar field's interaction with gravity provides its action. A fictitious field called the scalar field has been proposed to explain a variety of physics phenomena, including inflation, DE, and the Higgs technique. The action tries to generalize the theory of GR using basic concepts of scalar-tensor gravitational theories.  In this situation, the scalar field is vital in adjusting the gravitational force's strength, which results in a wide range of physical events. Usually, for a scalar field, the action is expressed in terms of the scalar field and its derivatives as
%The action is a basic idea of scalar-tensor theories of gravity which attempt to generalize the concepts of GR theory. In this situation, the scalar field is vital in adjusting the gravitational force's strength, which results in a wide range of physical events. Usually, the action of the scalar field is expressed in terms of scalar field and its derivatives as
\begin{equation}
S_{\phi} =  \int {d^4 x \sqrt{-g} \biggl[\frac{1}{2} \delta_{\nu}\phi\delta^{\nu}\phi  - V(\phi) \biggl] }
\end{equation}
where $\phi$ and $V(\phi)$ are scalar field and scalar potential respectively.\\
In the scalar field background, Klein-Gordon equation is read as
\begin{equation}
\frac{d^2 {\phi}(t)}{dt^2} + 3 \frac{d {\phi}(t) }{dt} H+\frac{d V(\phi)}{d\phi} = 0
%\ddot{\phi}(t)+3 H \dot{\phi}(t)+\frac{d V(\phi)}{d\phi} = 0
\end{equation}
For the scalar field, the energy-momentum tensor is developed as $T^{\phi}_{ij} = (\rho_{\phi}+p_{\phi})u_{i} u_{j}-p_{\phi} g_{ij}$.
The pressure $p_{\phi}$ and energy density $\rho_{\phi}$ for the scalar field are expressed as \cite{ref38,ref39}
\begin{equation}
p_{\phi} = \frac{1}{2} \dot{\phi}^2 - V(\phi)
\end{equation}
\begin{equation}
\rho_{\phi} = \frac{1}{2} \dot{\phi}^2 + V(\phi)
\end{equation}
%Here, potential energy $V(\phi)$ and kinetic energy $\frac{\dot{\phi}^2}{2}$ are the functions of scalar field $\phi$. \\
where,  $V(\phi)$ and $\frac{\dot{\phi}^2}{2}$ respectively represent the potential and kinetic energies and  both are functions of $\phi$. \\
%The scalar field and gravitational field are frequently coupled using the scalar curvature coupling, which couples the Ricci curvature scalar to the scalar field. The Einstein equations change when this term affects the gravitational constant. The coupled action can be stated as follows:
The scalar field is frequently coupled with gravitational field using the coupling of scalar curvature, which connect the Ricci curvature to the scalar field. The Einstein equations change when this term affects the gravitational constant. The action due to coupling can be stated as follows:
\begin{eqnarray}
A & = & S_{\psi}+S_{\phi}\nonumber\\
& = & \int {d^4 x \sqrt{-g} \bigg[ \frac{1}{2} \psi^4 R +\mathcal{L}_{m} +\bigg(\frac{1}{2} \delta_{\nu}\phi\delta^{\nu}\phi  - V(\phi) \bigg)\bigg]}
\end{eqnarray}
Thus, the Friedmann field equations considering matter and scalar field as the source may be expressed as\cite{ref39a} 
%\begin{equation}
%3 H^2 -\frac{3}{4} \beta^2 =\rho_{eff} = \rho_{m}+\rho_{\phi}+\rho_{\beta}
%\end{equation}
%\begin{equation}
%2\dot{H}+3 H^2 +\frac{3}{4} \beta^2 = p_{eff}=-(p_{m}+p_{\phi}+p_{\beta})
%\end{equation}
\begin{equation}
3 H^2 =\rho_{eff} = \rho_{m}+\rho_{\phi}+\rho_{\beta}
\end{equation}
\begin{equation}
2\dot{H}+3 H^2  = p_{eff}=-(p_{m}+p_{\phi}+p_{\beta})
\end{equation}
where $p_{\beta} = \rho_{\beta} = \frac{3}{4} \beta^2$. Here  $p_{\beta}$ and $\rho_{\beta}$ are pressure and energy density due to displacement vector $\beta$. It is crucial to keep in mind that for the stiff fluid $p_{\beta} = \rho_{\beta}$. Our universe has already experienced a time when the pressure and matter density were equal. Assuming the universe to be dust filled ($p_{m}=0$), the equations of energy conservation for scalar field and matter are established as  
\begin{equation}
\frac{d\rho_{\phi}}{dt}+3 (p_{\phi}+\rho_{\phi}) H = 0
\end{equation}
\begin{equation}
\frac{d}{dt}\rho_{m}+3 \rho_{m} H = 0
\end{equation}
From Eqs. (13) and (14), we get the following solutions
\begin{equation}
\omega_{\phi} = -\bigg[1+\frac{1}{3 H}\bigg(\frac{\dot{\rho_{\phi}}}{\rho_{\phi}}\bigg)\bigg]
\end{equation}
where $\omega_{\phi}=\frac{p_{\phi}}{\rho_{\phi}}$ denotes the scalar field's EoS parameter.
\begin{equation}
\rho_{m} = \rho_{m0} \frac{1}{a^3}
\end{equation} 
where integrating constant $\rho_{m0}$ is referred as the present value of matter energy density. \\

In general, it is difficult to solve the system of Eqs. (11) and (12) involving $\rho_{\phi}$,  $\rho_{\beta}$, $p_{\phi}$, and $H $ as unknowns. Thus, we need some more variable or parametrized relations to provide a solution to the system. It is crucial to investigate cosmic models other than the cosmological constant because it is insufficient to fully explain the universe's accelerated expansion. Although various physical explanations are mentioned for choosing such constraints, model-independent approaches are well-known choices that are based on the specific parametrization of the EoS parameter, deceleration parameter, and energy density \cite{ref40,ref40a}. In the case of DE cosmic models, the EoS parameter expresses a relation between energy density and pressure.  The model with the cosmological constant is assumed as the standard DE model where the EoS parameter remains constant and its value is found as -1.  However, the study of variable EoS parameters provides information on the underlying physics of DE. A simplest two-parameter model known as Chevallier-Polarski-Linder (CPL) parametrization has the potential to detect variations from a fixed EoS value\cite{ref41,ref42}. To examine DE possibilities beyond the cosmological constant, parametrizations that are considerably more complex can also be utilized, including the Jassal-Bagla-Padmanabhan (JBP) \cite{ref43,ref44}, the hybrid \cite{ref45}, and the BA \cite{ref46}. Another approach is to parametrize the DE energy density as a cosmic time t (or, alternatively, redshift z) function. Polynomial expansions and principal component analysis are two techniques that can be used to achieve this \cite{ref45,ref47,ref48,ref49}. These approaches can provide insight into how DE behaved over different cosmic eras. Here, the scalar field's energy density is considered as the source of dark energy and suitably parametrized in the following form
\begin{equation}
\rho_{\phi}=\rho_{\phi 0} (1+z)^{\alpha} e^{n z}
\end{equation}
where $\alpha$ and $n$ are constants, and $\rho_{\phi 0}$ is the present value of universe's critical density. These model parameters will be constrained from observational datasets. In the background of scalar fields theory, several cosmological models and high energy theories are explored using hybrid energy density as an explicit choice \cite{ref45}. In the present study, our focus is to utilize the hybrid parametrization of the potential as a phenomenological method to examine the evolutionary behavior of the cosmos in the framework of scalar-tensor theory.
%where $\alpha$ and $n$ are constants, and $\rho_{\phi 0}$ is the present value of universe's critical density. These model parameters will be constrained from observational datasets. In background of scalar fields theory, several cosmological models and high energy theories are explored using hybrid energy density as an explicit choice \cite{ref45}. In the present study, our focus is to utilize the hybrid parametrization of the potential as a phenomenological method to examine the evolutionary behaviour of cosmos in framework of scalar tensor theory.
It is crucial to emphasize that our goal is not to describe certain high-energy theories of physics that anticipate the specific form of the hybrid potential mentioned in Eq. (17). We instead use this parametrization as a phenomenological technique to study the behaviour and implications of the scalar field dark energy hypothesis.

% Instead to describe a certain high-energy theories of physics which anticipates the specific form of the hybrid potential mentioned in Eq. (17). It is a technique for examining the model's consistency with empirical data and for gaining understanding of the dynamics of DE.

%It is crucial to emphasise that our goal is not to assert that a particular high-energy physics theory can accurately predict the shape of the hybrid potential employed in Eq. (17). We instead use this parametrization as a phenomenological method to examine the actions and effects of the scalar field dark energy hypothesis. It is a method for investigating how well the model fits with observational data and for gaining understanding of the dynamics of DE.
For redshift transformation, we can utilize the relation $a = \frac{a_{0}}{1+z}$, where  $a$ is the average value of scale factor and present value $a_{0}$ is assumed to be 1. 
%In terms of redshift z, the scale factor can be realised as $a = \frac{a_{0}}{1+z}$. In our study, the present value of scale factor $a_{0}$ is taken as 1. The Hubble parameter in terms of z can be developed as $H = -(1+z) \frac{d H}{dz}$.
Thus, using Eq.(16), the matter energy density in terms of redshift z can be calculated as
\begin{equation}
\rho_{m} =  (1+z)^3 \, \rho_{m0}
\end{equation} 
Now, from Eqs. (17), (18), and (11), we get
\begin{equation}
3 H^2 = (1+z)^3 \, \rho_{m0} + \rho_{\phi 0} (1+z)^{\alpha} e^{n z} +\rho_{\beta}
\end{equation}
Using the gauge function $\rho_{\beta} = \beta_{0} a^{-2 k}$, Thus, Eq.(19) can be recast as 
\begin{equation}
H(z) = H_{0} \sqrt{(1+z)^3 \, \Omega_{m0} + (1+z)^{\alpha} e^{n z} \, \Omega_{\phi 0} +\Omega_{k0} (1+z)^{2 k}}
\end{equation}
where $\Omega_{m} = \frac{\rho_{m}}{\rho_{c}}$, $\Omega_{\phi} = \frac{\rho_{\phi}}{\rho_{c}}$, and $\Omega_{\beta} = \frac{\rho_{\beta} }{\rho_{c}}$ are the unit-less density parameters for the proposed model. These dimensionless parameters play a key role in explaining the whole content of the cosmos. The $\rho_{c} = 3 H^2$ is the critical density of the universe.
Here, $H_{0}$ denotes the present value of Hubble constant, subscripted $\Omega_{i 0}$ represents the values of density parameters at $z=0$. Thus, Eq.(2) can be reduced as follows for $z=0$. 
\begin{equation}
\Omega_{m0}+\Omega_{\phi 0}+\Omega_{k0} = 1
\end{equation} 
%The expression for deceleration parameter $q$ is given by
%\begin{eqnarray}
%q &=& -1 + \frac{1}{H(z)} (1+z) \frac{dH}{dz} \nonumber\\
%&=& \frac{2 (k-1) (z+1)^{2 k} \Omega _{\text{k0}}+(z+1)^3 \Omega _{\text{m0}}+e^{n z} (z+1)^{\alpha } \Omega _{\text{$\phi $0}} (\alpha +n z+n-2)}{2 \left((z+1)^{2 k} \Omega _{\text{k0}}+(z+1)^3 \Omega _{\text{m0}}+e^{n z} (z+1)^{\alpha } \Omega _{\text{$\phi $0}}\right)}
%\end{eqnarray}
From Eq. (15), the EoS parameter of scalar field can be derived as
\begin{equation}
\omega_{\phi} = \frac{1}{3} (\alpha +n z+n-3)
\end{equation}
So, for the proposed model the effective EoS parameter can be read as 
\begin{eqnarray}
\omega_{eff} &=& \frac{p_{eff}}{\rho_{eff}}=\frac{p_{\phi}+p_{\beta}}{\rho_{m}+\rho_{\phi}+\rho_{\beta}}\\
&=&\frac{(2 k-3) \left(1-\Omega_{m0}-\Omega_{\phi 0}\right)  (z+1)^{2 k}+\Omega_{\phi 0} e^{n z}  (\alpha +n z+n-3) (z+1)^{\alpha }}{3 \bigg[\left(1-\Omega_{m0}-\Omega_{\phi 0}\right) (z+1)^{2 k}+ \Omega_{m0} (z+1)^3+e^{n z} (z+1)^{\alpha } \Omega _{\phi 0}\bigg]}\nonumber
\end{eqnarray}
Now, for the proposed model the expressions of density parameters for matter, scalar field, and Lyra factor respectively, are given in the following forms.
\begin{equation}
\Omega_{m} (z) = \frac{\rho_{m}}{3 H^2} = \frac{(z+1)^3 \Omega _{m0}}{\left(-\Omega_{m0}-\Omega_{\phi 0}+1\right) (z+1)^{2 k}+\Omega_{m0} (z+1)^3+e^{n z} (z+1)^{\alpha } \Omega_{\phi 0}}
\end{equation}
\begin{equation}
\Omega_{\phi} (z) = \frac{\rho_{\phi}}{3 H^2} = \frac{e^{n z} (z+1)^{\alpha } \Omega _{\phi 0}}{(z+1)^{2 k} \left(-\Omega _{m0}-\Omega _{\phi 0}+1\right)+(z+1)^3 \Omega _{m0}+e^{n z} (z+1)^{\alpha } \Omega _{\phi 0}}
\end{equation}
\begin{equation}
\Omega_{\beta} (z) = \frac{\rho_{\beta}}{3 H^2} = \frac{(z+1)^{2 k} \left(-\Omega_{m0}-\Omega _{\phi 0}+1\right)}{(z+1)^{2 k} \left(-\Omega _{m0}-\Omega _{\phi 0}+1\right)+(z+1)^3 \Omega _{m0}+e^{n z} (z+1)^{\alpha } \Omega _{\phi 0}}
\end{equation}
From Eqs. (8), (9), (13), and (14), the kinetic and potential energies of the scalar field are given through the following expressions. 
%\begin{eqnarray}
\begin{equation}
\frac{\dot{\phi}^2}{2} =  -\frac{1}{2} H_0^2 \bigg[(2 k-3) \left(\Omega_{m0}+\Omega_{\phi0}-1\right) (z+1)^{2 k}- e^{n z} \Omega_{\phi0} (z+1)^{\alpha} (\alpha +n z+n)\bigg]
\end{equation}
%\end{eqnarray}
%\begin{eqnarray}
\begin{equation}
V(\phi) = \frac{1}{2} H_0^2 \bigg[(2 k-3) \left(\Omega_{m0}+\Omega_{\phi0}-1\right) (z+1)^{2 k}- e^{n z} \Omega_{\phi0} (z+1)^{\alpha} (\alpha +n z+n-6)\bigg]
\end{equation}
%\end{eqnarray}
The parameters ($H_{0}$, $\Omega_{m0}$, $\Omega_{\phi0}$, $\alpha$, $n$, $k$) have a significant impact on the model presented in Eq. (20), which determine the behavior and cosmological characteristics of the model. In the next segment, our aim is to analyze current experimental data to better recognize the significance of the present model. We specifically intend to study how the behavior of cosmological parameters is affected by constraining the values of important parameters ($H_{0}$, $\Omega_{m0}$, $\Omega_{\phi0}$, $\alpha$, $n$, $k$).
\section{Datasets and Cosmological Constraints Methods}
\subsection*{Supernovae type Ia}
%We use the Pantheon compilation sample consisting 1048 data point in the redshift range $0.01\leq z \leq 2.26$ [].This sample contains 276 SNIa case from PanSTARRSI Medium Deep Survey, Low-$z$ and HST samples. We use the systematic covariance $C_{sys}$ for a vector of binned distances
We have considered the sample consisting of 1048 points of Pantheon compilation with redshifts between $0.01$ and $2.26$. 276 SNIa points from the PanSTARRSI Medium Deep Survey, Low-$z$, and HST samples are included in this sample \cite{ref50,ref51}. 
%For a vector of binned distances, we utilize the systematic covariance $C_sys$.
%\begin{equation}
%C_{ij, sys} = \sum_{n = 1}^{i} {\left(\frac{\partial \mu_{i}}{\partial S_{n}}\right)\left(\frac{\partial \mu_{j}}{\partial S_{n}}\right) \sigma_{S_{k}}}
%\end{equation}
%in which the summation is over the n systematic with $S_n$ and magnitude of its error $\sigma_{S_{k}}$. 
For the sample of Pantheon predictions, the $\chi^2$ measure is defined by the following relation. 
\begin{equation} 
\chi^{2}_{SN}=(\mu_{obs}-\mu_{th})^{T}\left(C^{-1}_{SN}\right)(\mu_{obs}-\mu_{th})
\end{equation}
where $\mu_{th}= 5\log_{10}{\frac{c DL}{H_{0}Mpc}}+25$, $\mu_{obs}$ be the observed distance modulus, and for the Pantheon sample, $C_{SN}$ denotes the covariance matrix \cite{ref50}. $H_{0}$ indicates the Hubble rate while $c$ corresponds to speed for a particle of light. For a flat FRW universe, the luminosity distance is expressed as $D_{L}(z)=(1+z)H_{0} \int_{0}^{z} \frac{dx'}{H(x')}$. To limit the parameters of the proposed model for the Pantheon compilation sample, we use the following statistical measure.
\begin{equation}
\chi^{2}_{Pantheon} = \Delta \mu^{T}. C^{-1}_{Pantheon}. \Delta \mu
\end{equation} 
in which $\Delta \mu = \mu_{data}-\mu_{obs}-M$ and $M$ corresponds to a nuisance parameter. The entire collection of full and binned Pantheon supernova data is available online \cite{ref52}.

%It should be noted that the $C_{Pantheon}$ is the summation of the systematic covariance and statistical matrix $D_{stat}$ having a diagonal component. The complete version of full and binned Pantheon supernova data are provided in the online source [].

\section*{BAO/CMB data} 
%We have considered BAO [] and CMB[] measurements dataset to obtain the BAO/CMB constraints on the model parameters. We have considered six BAO/CMB data points (see Table 1). For BAO dataset, the results from the WiggleZ Survey[], SDSS DR7 Galaxy sample[], and 6dF Galaxy Survey[] datasets have been used. On the other hand, the CMB measurement considered is derived from the WAMP7 observations[]. The discussion as regards the BAO/CMB dataset has also been presented in very similar way in [], but details of methodology for obtaining the BAO/CMB constraints on model parameters is available in Ref.[]. For this dataset, the $\chi^2$ function is defined as
To determine the restrictions on parameters of the model, we took into account the BAO \cite{ref53,ref54,ref55} and CMB \cite{ref56,ref57} measurements dataset. Six BAO/CMB data points have been considered (Table 1). For the BAO sample, the predictions from a sample of Galaxy Surveys like SDSS DR7 and 6dF, and WiggleZ have been utilized\cite{ref53,ref54,ref55}. However, the CMB measurement under consideration is based on WAMP7 observations \cite{ref56}. A similar explanation of the given sample can be seen in \cite{ref45,ref58}, but \cite{ref58} provides more information on the approach used and sample to constrain the parameters.

%Results from the WiggleZ Survey, SDSS DR7 Galaxy sample, and 6dF Galaxy Survey datasets have all been used for the BAO dataset.
%The explanation of the BAO/CMB dataset is also presented in a very similar manner in [], but [] provides more information on the approach used to derive the BAO/CMB constraints on model parameters. 

The angular diameter distance for the sample is defined as $D_{A}=\frac{D_{L}}{(1+z)2}$, where $D_{L}$ indicates the proper angular diameter distance \cite{ref58}, and the dilation scale is described by $D_{V}(z)=\left[ D^{2}_{L}(z)*(1+z)^2*\frac{c \,  z}{H(z)} \right]^{1/3}$. \\

%$D_{H}(z)/r_{d} = c/{H(z) r_{d}}$, $r_d$ represents the sound horizon at the drag epoch, 

%%%%%%%%%%%%%%%%%%%%%%%%%%%%%%% Table %%%%%%%%%%%%%%%%%%%%%%%%%%%%%%%%%%%%%%%%
\begin{table}
	%	\caption{ The best fit values of free parameters for Model-1 for different observational dataset }
	\begin{center}
		\begin{tabular}{|c|c|c|c|c|c|c|}
			\hline 
			\multicolumn{7}{|c|}{ Values of $\varUpsilon(z)$ for different points of $z_{BAO}$}\\
			\hline
			$z_{BAO}$& $0.106$ & $0.2$	 & $0.35$  & $0.44$ & $0.6$ &$ 0.73$  \\
			\hline
			\small  $\varUpsilon(z)$ & \small$30.95 \pm 1.46$ & \small$17.55 \pm 0.60$  & \small$10.11 \pm 0.37$ & \small $8.44 \pm 0.67$ &\small $6.69 \pm 0.33$ & $ 5.45 \pm 0.31$\\
			\hline
		\end{tabular}
	\end{center}
\end{table}
Here, $\varUpsilon(z)= d_A(z_{*})/D_V(z_{BAO})$ and $z_{*}\approx 1091$.\\
For limiting the parameters of the model,  the chi-square estimator for the BAO sample is described in the following form \cite{ref45,ref58,ref59,ref59a}.
\begin{equation}
\chi^2_{BAO/CMB} = X^{T} C^{-1} X
\end{equation}

where 
\begin{align*}
X &= \begin{pmatrix}
\frac{d_A(z_{*})}{D_V(0.106)}-30.95\\           
\frac{d_A(z_{*})}{D_V(0.20)}-17.55\\
\frac{d_A(z_{*})}{D_V(0.35)}-10.11\\
\frac{d_A(z_{*})}{D_V(0.44)}-8.44\\
\frac{d_A(z_{*})}{D_V(0.60)}-6.69\\
\frac{d_A(z_{*})}{D_V(0.73)}-5.45
\end{pmatrix}
\end{align*}
and $C^{-1}$ is given by \cite{ref58}
\begin{align*}\label{15}
C^{-1} &= \begin{pmatrix}
0.48435 & -0.101383 & -0.164945 & -0.0305703 &-0.097874 & -0.106738\\           
-0.101383 & 3.2882 &-2.45497 &-.0787898 &-0.252254 &-0.2751\\
-0.164945 &-2.454987 &9.55916 &-0.128187 &-0.410404 &-0.447574\\
-0.0305703 &-0.0787898 &-0.128187 &2.78728&-2.75632 &1.16437\\
-0.097874 &-0.252254 &-0.410404 &-2.75632 &14.9245 &-7.32441\\
-0.106738 &-0.2751 &-0.447574 &1.16437 &-7.32441 &14.5022
\end{pmatrix}.
\end{align*}

\section*{Observational Hubble Data (OHD)}
%We have take over 57 $H(z)$ datapoints for z ranging in between $0.07$ and $2.36$ calculated from cosmic chronometric technique, galaxy clusters[], and differential age technique, which additionally yield the redshift dependence of the Hubble parameter as $(1+z) H(z) = -\frac{dz}{dt}$. The chi-square function is used to compare the theoretical predictions ($E_{th}$) of the model with observations ($E_{obs}$). 
We have take-over 57 $H(z)$ datapoints for z ranging in between $0.07$ and $2.36$ calculated from cosmic chronometric technique, galaxy clusters \cite{ref32}, and differential age procedure. Then the Hubble constant can be realized in the form of redshift as $(1+z) H(z) = -\frac{dz}{dt}$.  Now, the estimator $\chi^2$ is taken into consideration for the purpose of limiting the model’s parameters by comparing the model’s theoretical predictions ($E_{th}$) with experimental values ($E_{obs}$).`
\begin{equation}
\chi^{2}_{OHD} = \sum_{i= 1}^{57} {\frac{\big[E_{th}(z_i)-E_{obs}(z_i)\big]^2}{\sigma^2_i}}
\end{equation}
where $\sigma_{i}$ is the error detected in experimental estimations of $H(z)$.\\
%Hence, the total $\chi^2$ for the combined dataset (SNIa+BAO/CMB+OHD) is given by
Thus, the joint estimator for a combined sample of experimental predictions including BAO/CMB, OHD, and SNIa samples, the combined statistic measure is defined in the following manner\cite{ref45,ref58,ref59,ref59a}.
\begin{equation}
\chi^2_{tot} = \chi^2_{Pantheon}+\chi^2_{BAo/CMB} + \chi^{2}_{OHD}
\end{equation}
The $\chi^{2}_{tot}$  statistic can be minimized to find the parameter value that best fits the combined sample of the SNIa, OHD, and BAO/CMB datasets.  By taking maximum likelihood approach into account, the total likelihood function $\mathcal{L}_{tot} = exp(-\chi^2_{tot}/2)$ may be calculated as the product of individual likelihood functions of each dataset expressed in the form $\mathcal{L}_{tot}= \mathcal{L}_{Pantheon} * \mathcal{L}_{BAO/CMB}* \mathcal{L}_{OHD}$. The likelihood function $\mathcal{L}_{tot}(x*)$ is maximized or, alternatively $\chi^2_{tot} (x^{*})=-2 \ln \mathcal{L}_{tot} (x^{*})$ is minimized to get the most plausible values of parameters. For the set of cosmic parameters (pointed at $x^{*}$), the $1 \sigma$ and $2 \sigma$ contours are constrained and bounded respectively by $\chi^2_{tot} (x)=\chi^2_{tot} (x^{*})+2.3$ and $\chi^2_{tot} (x)=\chi^2_{tot} (x^{*})+6.17$.  We get best-fit parameter values for the derived model by minimizing the $\chi^2$ statistic. 
%%%%%%%%%%%%%%%%%%%%%%%%%%%%%%%%% Fig 1 %%%%%%%%%%%%%%%%%%%%%%%%%%%%%%%%%%%%%%%
\begin{figure}
\centering
\includegraphics[scale=0.45]{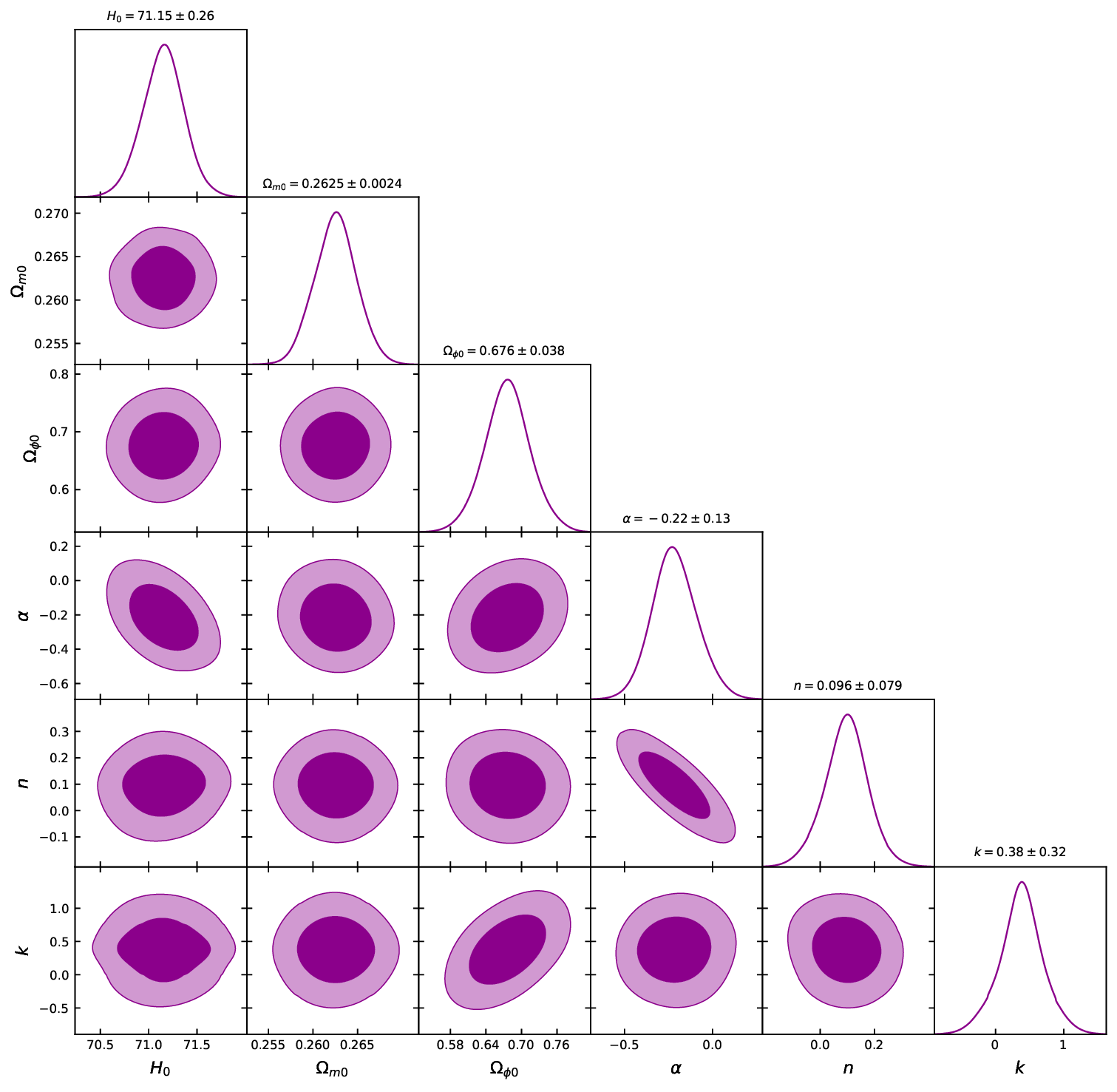}
%(b)\includegraphics[width=7cm,height=5cm,angle=0]{fig3b.eps}
\caption{Confidence contour plot for joint data set of OHD, Pantheon and BAO.} 
\end{figure}
%%%%%%%%%%%%%%%%%%%%%%%%%%%%%%%%%%%%%%%%%%%%%%%%%%%%%%%%%%%%%%%%%%%%%%%%
%%%%%%%%%%%%%%%%%%%%%%%%%%%%%%%%% Fig 2 %%%%%%%%%%%%%%%%%%%%%%%%%%%%%%%%%%%%%%%
\begin{figure}
\centering
(a)\includegraphics[width=5.6cm,height=5.0cm,angle=0]{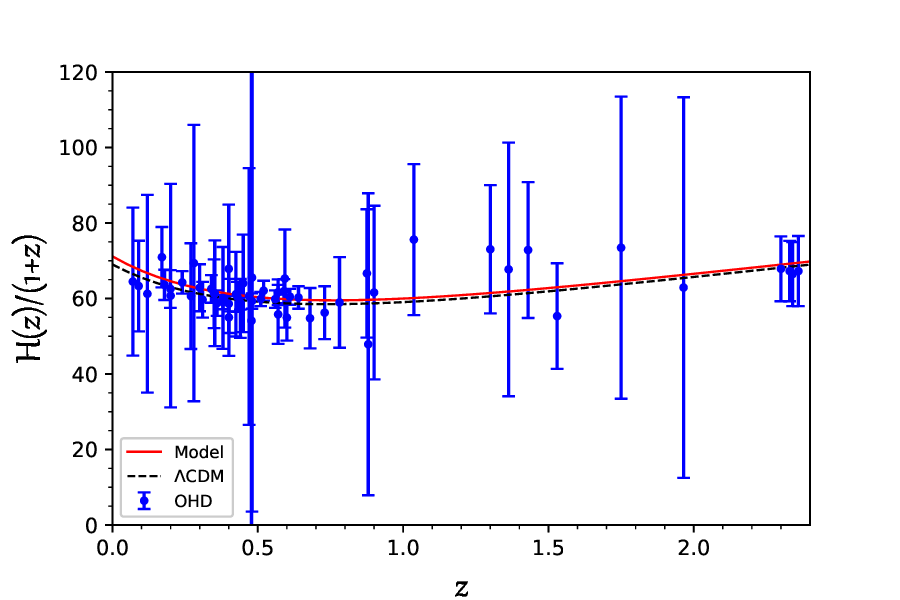} 
(b)\includegraphics[width=5.6cm,height=5.0cm,angle=0]{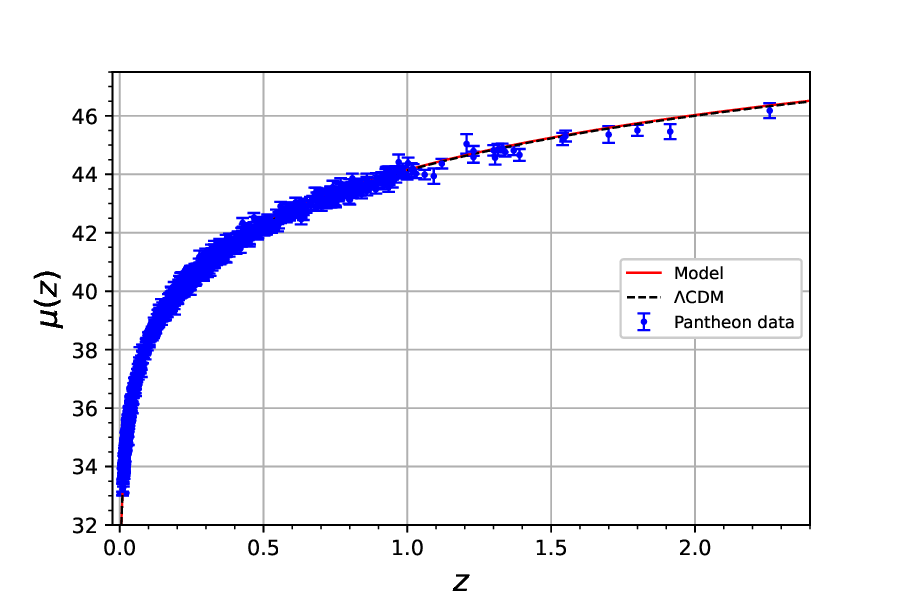}
\caption{(a) Plot of Hubble rate $H(z)/(1+z)$ vs $z$ for $ H_0 = 71.15\pm 0.26$ km/s/Mpc , (b) Distance modulus plot versus $z$ for $ H_0 = 71.15\pm 0.26$ km/s/Mpc.}
\end{figure}
%%%%%%%%%%%%%%%%%%%%%%%%%%%%%%%%%%%%%%%%%%%%%%%%%%%%%%%%%%%%%%%%%%%%%%%%%%%%%%%
Figure 1 displays the statistical results in confidence contours with $1\sigma$ and $2\sigma$ limits for the proposed model utilizing the joint dataset of SN, BAO/CMB, and OHD. The best plausible values of parameters estimated from the joint dataset are summarized in Table 1. \\
The comparative behavior of the suggested model with the existing standard models and relevant datasets is plotted and presented in Fig.2. The Hubble rate ( $H(z)/(1+z)$) as a function of $z$ is plotted for the purpose. In Figure 2(a), the solid red line depicts the behaviour of our suggested model utilizing the values of parameters obtained from the joint dataset, the 57 data points of OHD are represented by error bars of blue colour, and the dashed black line denotes the findings of the traditional $\Lambda$CDM model. For a similar reason, Figure 2(b) plots the distance modulus $mu(z)$ as a function of $z$. For an observed supernova the difference between the ``apparent and absolute magnitude describes the luminosity distance which can be expressed as $ dL = a_{0} (1+z) r = (1+z) \int_{0}^{z}\frac{dz}{H(z)} $. This distance parameter, $\mu (z)$ is numerically equal to $25 + 5log10(dL/Mpc)$. Here in Figure 2(b), the blue error bar shows the points of the SN data that were taken into consideration as discussed earlier, the dashed black line shows the output of the traditional $\Lambda$CDM model, and the solid red line illustrates the characteristics of our derived model for the joint dataset. In our analysis, for the joint dataset, the estimated values of parameters are $ H_0 = 71.15\pm 0.26$ km/s/Mpc, $ \Omega_{m0}=0.2625\pm 0.0024$, $ \Omega_{\phi0} = 0.676\pm0.038$. These findings are consistent with the most recent results \cite{ref60,ref61,ref61a,ref62a,ref62b,ref63,ref64,ref64a,ref64b,ref64c,ref64d}.

%%%%%%%%%%%%%%%%%%%%%%%%%%%%%%%%%%%%%%%% Table 1 %%%%%%%%%%%%%%%%%%%%%%%%%%%%%%%%%
\begin{table}
\caption{ The best-fit values model parameters for joint observational dataset }
	\begin{center}
		\begin{tabular}{|c|c|c|c|c|c|c|c|}
			\hline 
			\tiny Parameters & \tiny $H_{0}$ & \tiny $\Omega_{m0}$	& \tiny $\Omega_{\phi0}$ & \tiny $\Omega_{k0}=1-\Omega_{m0}-\Omega_{\phi0}$ & \tiny $\alpha$ & \tiny $n$ & \tiny $k$  \\
			\hline
			\tiny Priors & \tiny $(60,80)$ & \tiny $(0,1)$  & \tiny $(0,1)$ & \tiny - & \tiny $(-1,0)$ & \tiny $(0,1)$  & \tiny $(0,1)$ \\ 
			\hline
%			\tiny Pantheon & \tiny $73.8\pm1.7$ & \tiny $0.252\pm0.033$  & \tiny $0.672\pm0.040$ & \tiny $-1.094\pm0.089$ & \tiny $0.0046\pm0.0034$ & \tiny $0.0044\pm0.0033$   \\ 
%			\hline
%			\tiny BAO & \tiny $67.29\pm0.31$ & \tiny $0.2622\pm0.0021$  & \tiny $0.7293\pm0.008$ & \tiny $-1.41\pm0.12$ & \tiny $0.00013\pm0.00014$ & \tiny $0.00013\pm0.00013$   \\
%			\hline
			\tiny SN+BAO/CMB+OHD & \tiny $71.15\pm 0.26$ & \tiny $0.2625\pm 0.0024$  & \tiny $0.676\pm0.038$ & \tiny $0.0615\pm0.0404$ & \tiny $-0.22\pm0.13$ & \tiny $0.096\pm0.079$ &\tiny $0.38\pm0.32$  \\
			\hline
		\end{tabular}
	\end{center}
\end{table}
%%%%%%%%%%%%%%%%%%%%%%%%%%%%%%%%%%%%%%%%%%%%%%%%%%%%%%%%%%%%%%%%%%%%%%%%%%%%%%%%%%%%%%%%%%%%
\section{Evolutionary behaviour of the model} 
The deceleration parameter (DP) is a critical parameter that portrays the transition era of the universe among the other parameters which describe the universe's evolutionary dynamics. In relation to the Hubble parameter, the DP can be realized as $q = -\frac{\ddot{a}}{H^{2} a} = -1+ \frac{1}{H} (1+z) \frac{d}{dz} H(z)$. Hence, for the proposed model the deceleration parameter can be derived as:
\begin{eqnarray}
q &=& -1 + \frac{1}{H(z)} (1+z) \frac{dH}{dz} \nonumber\\
&=& \frac{2 (k-1) (z+1)^{2 k} \Omega _{k0}+(z+1)^3 \Omega _{m0}+e^{n z} (z+1)^{\alpha } \Omega _{\phi 0} (\alpha +n z+n-2)}{2 \left((z+1)^{2 k} \Omega _{k0}+(z+1)^3 \Omega_{m0}+e^{n z} (z+1)^{\alpha } \Omega_{\phi 0}\right)}
\end{eqnarray}

%Analysis of $q$ along with Hubble parameter $H$, explains most of the universe expansion features. Age as well as phase transitions of universe from accelerated to decelerated expansion or vice versa can be accurately described from $q$ and $H$.\\
The major evolutionary features of the expanding universe can be analyzed and explained by the study of the Hubble parameter in consort with DP. The study of these parameters can accurately predict the various cosmic features of the evolutionary universe like age and phase transition (deceleration to acceleration or vice versa). \\

The inflationary behavior of the model can be justified by the sign of the DP $q$. While a negative sign of q depicts the universe's current accelerated expansion, a positive sign of q corresponds to the universe's decelerated expansion. For a combined sample of SNIa, OHD, and BAO/CMB datasets, the trajectory of the deceleration parameter is plotted in Fig.4. The earlier universe is thought to evolve with deceleration dynamics due to dark energy domination as clearly seen in Fig.4. Here, we see that the Universe in the derived model is currently going through an accelerated scenario of expansion. Additionally, for the derived model a signature flipping is observed at $z_{t} = 0.756^{+0.005}_{-0.015}$ and the model evolves in the acceleration phase in late time. At $z=0$, the present value of DP is observed as
\begin{eqnarray}
q &=& \frac{2 (k-1) \Omega_{k0}+ \Omega_{m0}+\Omega_{\phi0} (\alpha+n-2)}{2 \left(\Omega_{k0}+\Omega_{m0}+\Omega_{\phi0}\right)}
\end{eqnarray}
From the above analysis, the present value of DP ($q$) is observed to be $-0.625^{+0.067}_{-0.085}$, which greatly agrees with the recent findings. It is interesting to notice that $\frac{dH}{dt}_{t=t_0} = 0$ for $q_0 = -1$, which predicts a rapid expansion of the universe and the largest value of the Hubble parameter. Therefore, the dynamics of the late-time evolution of the observed Universe may be described using the Universe in the derived model. The derived outcomes of our theoretical model are nicely matched with the recent experimental results \cite{ref19,ref45,ref52,ref58,ref59,ref59a,ref61,ref61a,ref62,ref62b,ref63,ref64}.
%%%%%%%%%%%%%%%%%%%%%%%%%%%%%%% Figure 4 %%%%%%%%%%%%%%%%%%%%%%%%%%%%%%%%%%%%%%%%%%%%%%%%
\begin{figure}
\centering
\includegraphics[scale=0.60]{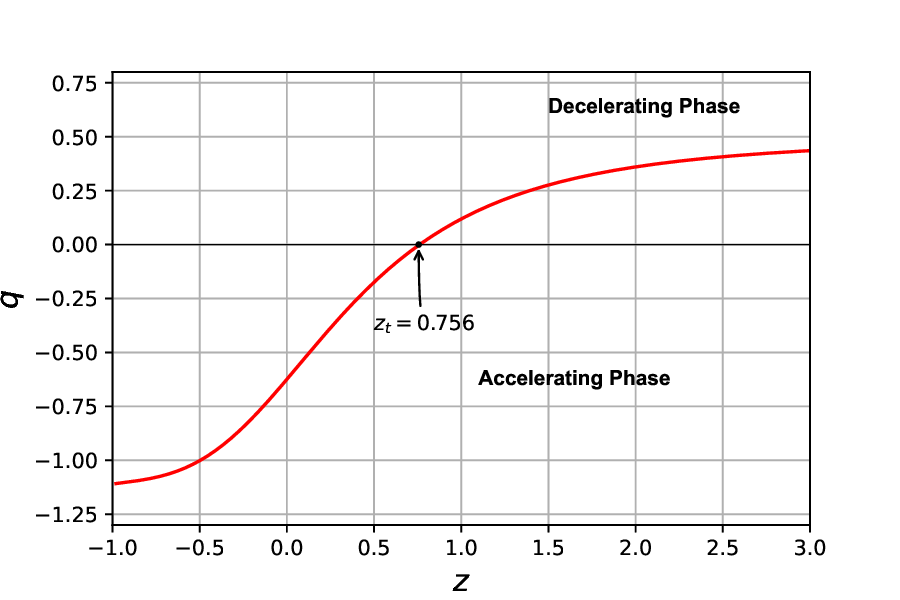}\label{fig4}
\caption{Deceleration parameter $q$ as a function of $z$ for OHD + Pantheon + BAO data sets.}
\end{figure}
%%%%%%%%%%%%%%%%%%%%%%%%%%%%%%%%%%%%%%%%%%%%%%%%%%%%%%%%%%%%%%%%%%%%%%%%%%%%%%%%%%%%%%%%%
The proposed model describes an expanding universe with the decrease in the densities of the scalar field and matter as shown in Fig. 3.  The density of the scalar field approaches to the least value in late time while the matter density advances to zero. Energy conservation in GR accounts for the drop in densities of the scalar field and matter with the expansion of the universe. The future evolution of the Universe will be significantly impacted by the assumption that in the late cosmos, the density of matter becomes zero. A phenomenon known as "heat death" is when the universe progresses gradually cold and dark, and no matter is left to create new galaxies or stars. The thermodynamics second rule, which stipulates that disorder or entropy can continuously increase over time, leads to this scenario. As the scalar field tends to have a smaller value of DE density in late time, the Universe can also keep expanding at an accelerating rate in the future. The scenario in which all the matter including stars and galaxies pulled apart, and the expansion of the universe grows very fast is recognized as the "big rip" scenario.\\
%%%%%%%%%%%%%%%%%%%%%%%%%%%%%%% Figure 4 %%%%%%%%%%%%%%%%%%%%%%%%%%%%%%%%%%%%%%%%%%%%%%%%
\begin{figure}
\centering
(a)\includegraphics[width=5.6cm,height=5cm,angle=0]{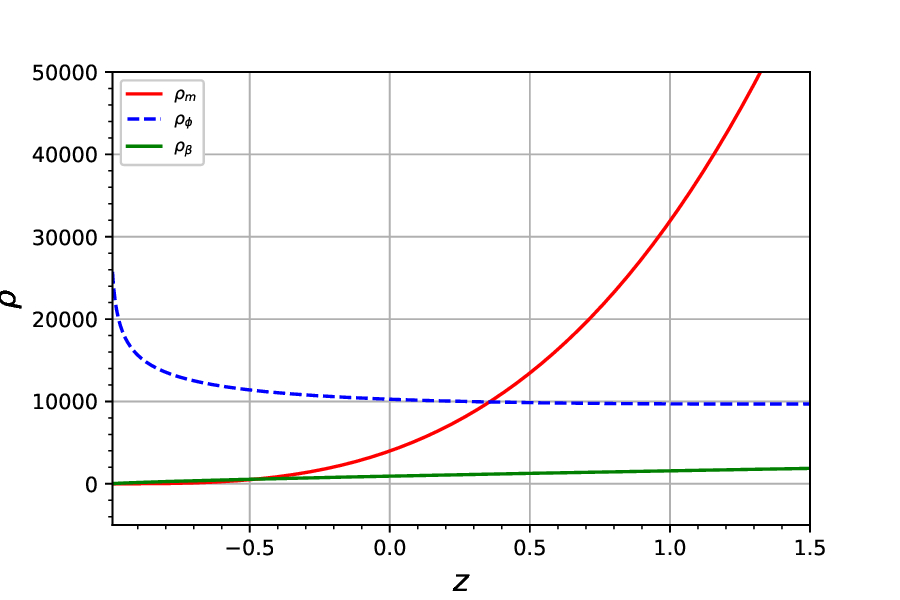}
(b)\includegraphics[width=5.6cm,height=5cm,angle=0]{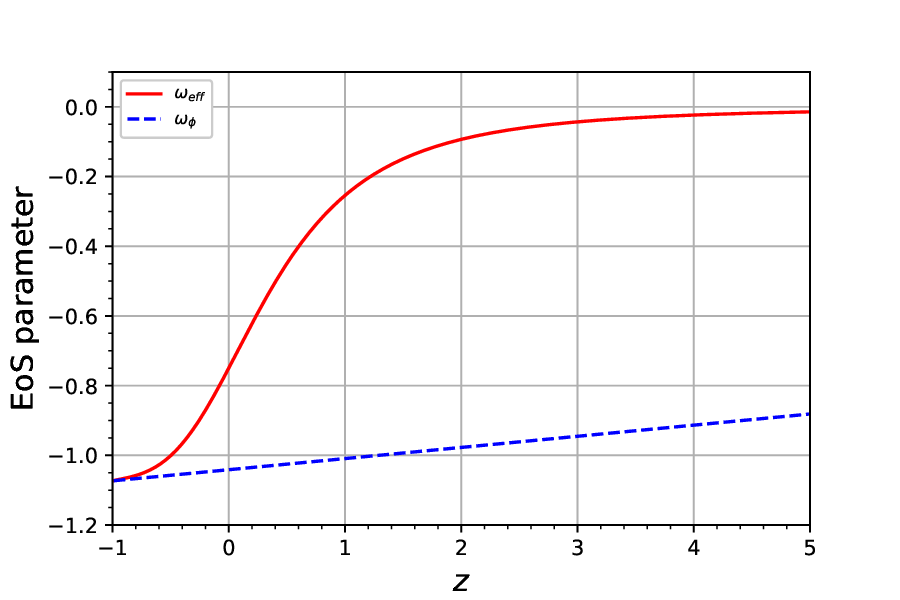}
\caption{(a) Energy density, (b) EoS.}
\end{figure}
%%%%%%%%%%%%%%%%%%%%%%%%%%%%%%%%%%%%%%%%%%%%%%%%%%%%%%%%%%%%%%%%%%%%%%%%%%%%%%%%%%%%%%%%%
In the same direction, the EoS parameter is another useful tool, which explains the evolutionary dynamics of the cosmos in terms of its rate of expansion.  The EoS parameter ($\omega$) can be expressed by a relation of the cosmic fluids energy density and the pressure in the form $p = \omega \rho$.  Based on the nature of pressure, the EoS parameter can be characterized by different cosmic realities. Dark matter is an example of non-relativistic matter for which $\omega= 0$, while for relativistic matter like radiation $\omega= 1/3$.  The accelerated or decelerated expanding nature of the cosmos can be characterized by distinct value of $\omega$. The scenario of accelerated expansion of the cosmos can be categorized into different conceivable DE scenarios, which include (i) quintessence scenario ($-1<\omega< -1/3$), (ii) cosmological constant ($\omega = -1$), (iii) phantom scenario ($\omega<-1$). \\

We have focused on both the effective EoS parameter and the EoS parameter of the scalar field in our model. According to GR, the only prerequisite for inflation in the cosmos is that which results in the cosmos having repulsive energy and jerk. In our investigation, we have validated the quintom behavior of the model by establishing the current values of the scalar field EoS parameter. The current value of the scalar field EoS parameter is estimated as $\omega_{\phi 0} = -1.042$ for best-constrained values of model parameters for a combined sample of datasets. This conclusion supports prior research in the field and lends substantial support to the quintom behavior of DE \cite{ref45,ref65,ref66,ref67,ref68}. Additionally, we have displayed the reconstructed evolution history of the effective EoS parameter $\omega_{eff}$ for this model using the combined sample of SN+OHD+BAO/CMB datasets in Figure 4. From the figure, it has been noticed that the model does not suffer any kind of `future singularity' because, at low redshift $\omega_{eff}$ attains a negative value ($\omega_{eff} < -1$) while at high z, it approaches to zero. Thus, from the trajectory of $\omega_{eff}$, it has been detected that the model proceeds in the quintessence era during the evolution of the cosmos and approaches to a phantom scenario in late time. Hence, the profile of $\omega_{eff}$ depicts a quintom-like behavior of the cosmos.\\
%%%%%%%%%%%%%%%%%%%%%%%%%% Figure 4 %%%%%%%%%%%%%%%%%%%%%%%%%%%%%%%%%%%%%%%%%%%
\begin{figure}
\centering
\includegraphics[scale=0.60]{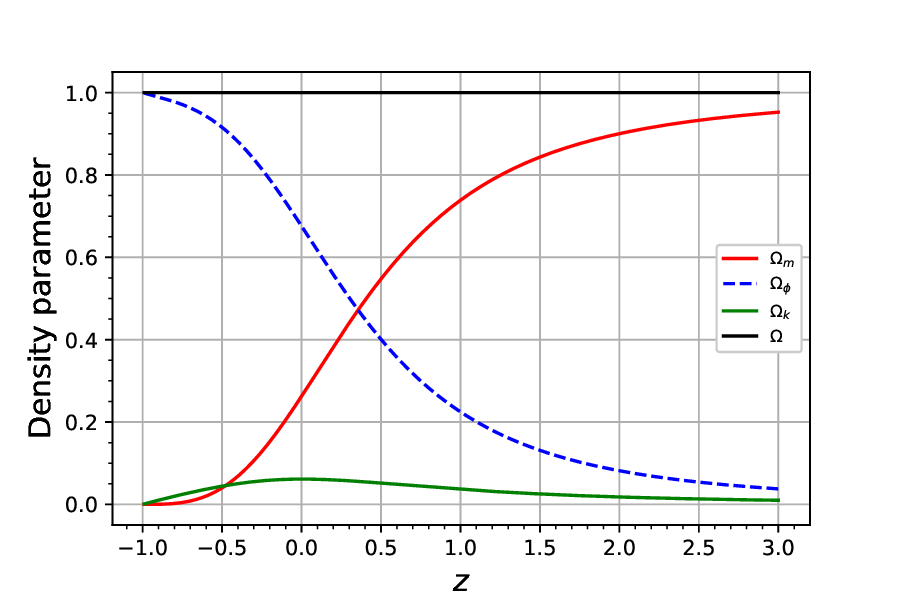}\label{fig4}
\caption{Deceleration parameter $q$ as a function of $z$ for OHD + Pantheon + BAO data sets.}
\end{figure}
%%%%%%%%%%%%%%%%%%%%%%%%%%%%%%%%%%%%%%%%%%%%%%%%%%%%%%%%%%%%%%%%%%%%%%%%%%%%%%%
Due to the dominance of non-relativistic matter like baryonic and dark matter in the early universe, the value of the scalar field's density parameter is low whereas the value of the matter density parameter is high i.e., $\Omega_{m}>\Omega_{\phi}$ which renders a strong physical background for the decelerated scenario of the early universe.  But, with the evolutionary growth of the universe, the density parameter decreases due to the volumetric increase of the universe and it tends to zero in late time whereas the density parameter of the scalar field becomes dominant in late time and leads to the universe’s accelerated expansion. The best-estimated values of density parameters of the proposed model for the combined sample of observation datasets (SN, OHD, and BAO) are found as $ \Omega_{m0}=0.2625\pm 0.0024$, $ \Omega_{\phi0} = 0.676\pm0.038$, and $\Omega_{k0}=0.0615\pm0.0404$. These outcomes are nicely matched with the findings of the Planck measurements \cite{ref34,ref69}. The total density parameter for the proposed model tends to unity in late time i.e., at the current era. Thus, in the present study, a scalar field is taken as a substitute for DE to describe the accelerated expansion of the cosmos. The development of the scalar field's kinetic and potential energy is also depicted in Figures 8 and 9. From the graphical representation of potential and kinetic energies, we observed that both are positive and decreasing, and the scalar field transit to a low energy scenario from a high energy era over time.
%%%%%%%%%%%%%%%%%%%%%%%%%%%%%%% Figure 4 %%%%%%%%%%%%%%%%%%%%%%%%%%%%%%%%%%%%%%%%%%%%%%%%
\begin{figure}
\centering
(a)\includegraphics[width=5.60cm,height=5cm,angle=0]{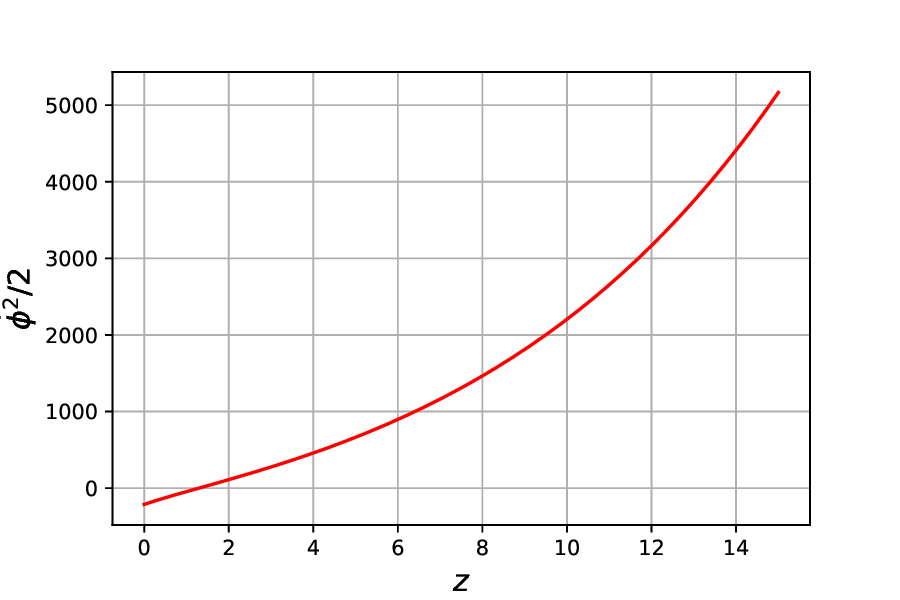}
(b)\includegraphics[width=5.60cm,height=5cm,angle=0]{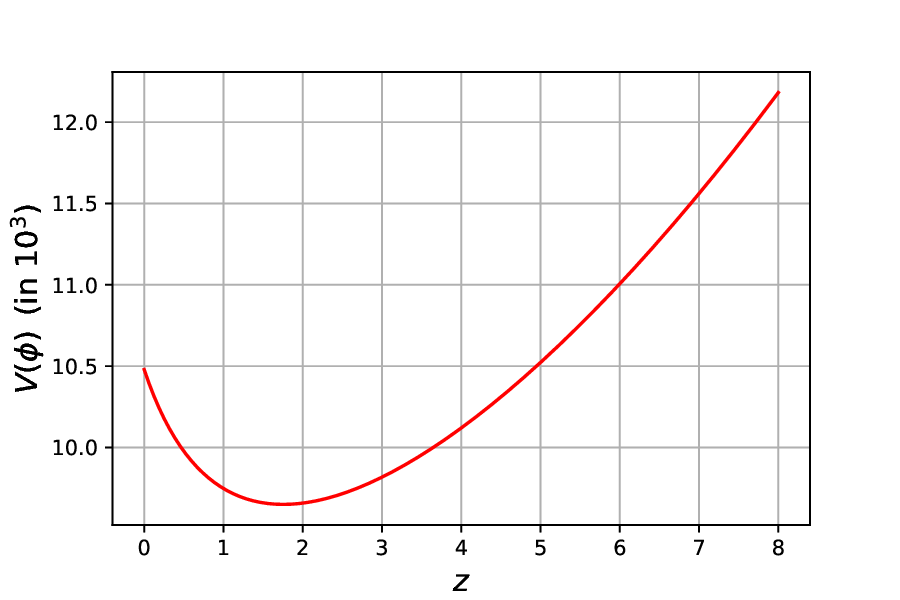}
\caption{(a) KE, (b) PE.}
\end{figure}
%%%%%%%%%%%%%%%%%%%%%%%%%%%%%%%%%%%%%%%%%%%%%%%%%%%%%%%%%%%%%%%%%%%%%%%%%%%%%%%%%%%%%%%%%
The energy associated with a scalar field, which has a single value at each space point, is described by the theory of the scalar potential in physics \cite{ref22,ref23,ref24}. The nature of the scalar field and its potential is highly influenced by the specific physical system taken into consideration. Obviously, a positive scalar potential is associated with stable structures in physics because negative energy can result in non-physical or unstable solutions. For the proposed model, the positive behavior of scalar potential during the entire evolution can be seen in Figure. The nature of kinetic energy is positive and decreasing in the evolution of the universe as depicted in the figure. Thus, the derived model shows a quintessence-like behavior in which massless scalar field $\phi$ can be assumed with a potential $V(\phi)$ that mimics the gravitational field and efficiently describes the current inflation of universe\cite{ref24,ref25,ref26}.
%%%%%%%%%%%%%%%%%%%%%%%%%%%%% Figure 4 %%%%%%%%%%%%%%%%%%%%%%%%%%%%%%%%%%%%%%%%
\begin{figure}
\centering
(a)\includegraphics[width=5.60cm,height=5cm,angle=0]{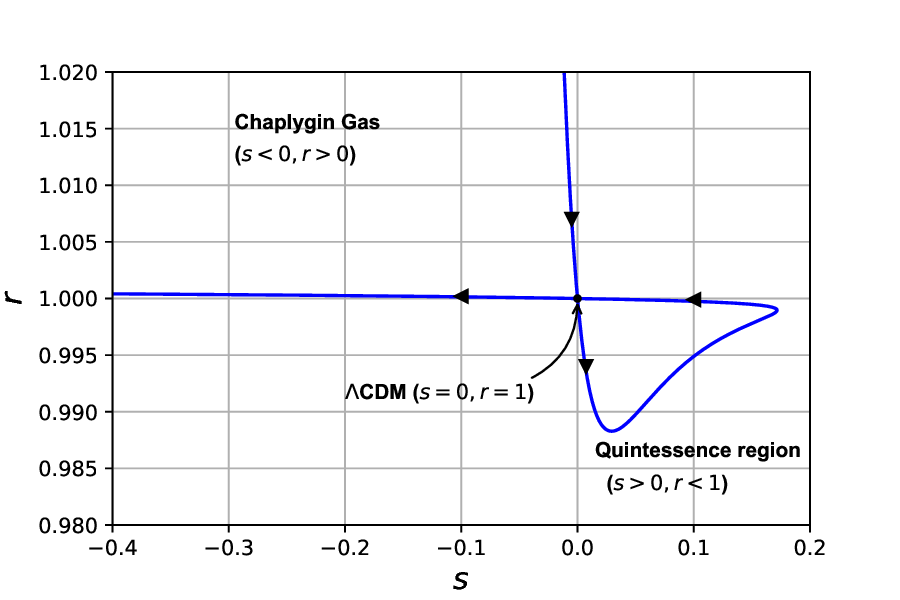}
(b)\includegraphics[width=5.60cm,height=5cm,angle=0]{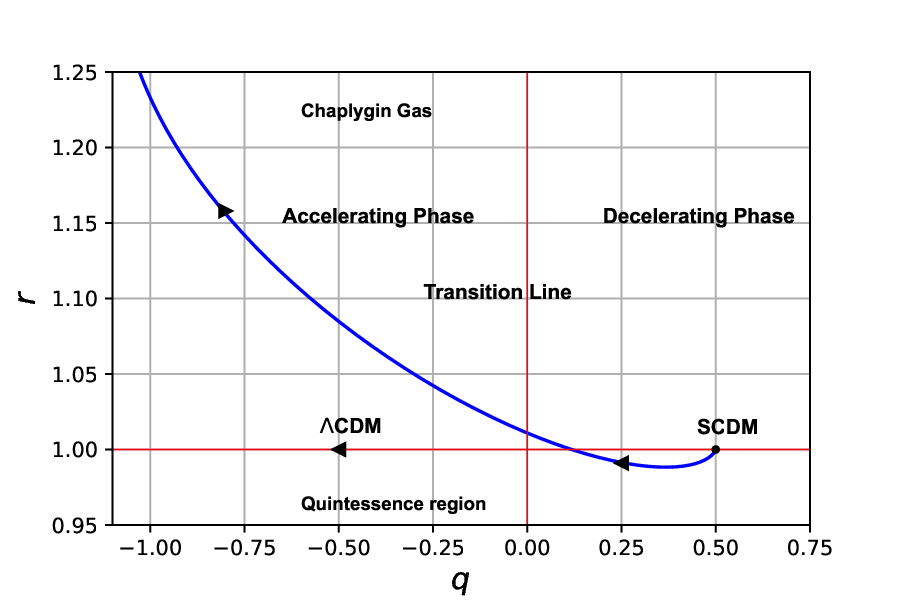}
\caption{(a)Trajectory in ($r-s$) plane, (b) )Trajectory in ($r-q$) plane.}
\end{figure}
%%%%%%%%%%%%%%%%%%%%%%%%%%%%%%%%%%%%%%%%%%%%%%%%%%%%%%%%%%%%%%%%%%%%%%%%%%%%%%%
\subsection{Statefinders}
Cosmological parameters like Hubble and deceleration are combined sufficiently to depict the evolutionary dynamics of the universe. Scale factor $a$, its first order derivative $(\dot{a})$, and second order derivative $(\ddot{a})$ are used to describe both of these parameters. The accuracy characteristics of the proposed theoretical models are suppressed as a result of this dependency, which causes all the models to converge around the same value of $q$ and other revealing parameters. Important theoretical model predictions about the accuracy of the outcomes are lost in the process. Two new parameters named as statefinder $r,s$ are presented to identify the degree of precision among various dark energy cosmic models.\\

This pair of state-finders assists us in enhancing model prediction accuracy by identifying evolutionary trajectory in the $r-s$ plane.  Assuming various forms of dark energies, as mentioned in the literature \cite{ref70,ref71,ref72}, it is possible to distinguish between the suggested cosmological model and the $\Lambda$CDM model from the $(r - s)$ plotting \cite{ref69a,ref70,ref73}.  In terms of $q$ and $z$, the parameters $r$ and $s$ for the currently suggested model can be elaborated as follows:
\begin{eqnarray}
r &=& q(2q+1)+(1+z) \frac{dq}{dz}\nonumber\\
&=&\bigg[2 (\Omega _{m0}+\Omega_{\phi 0}-1)(2 k^2 -3 k +1)(z+1)^{2 k} - 2(1+z)^3\Omega _{m0}+\Omega_{\phi 0} \bigg(2\nonumber\\
&-&e^{n z} (z+1)^{\alpha} \{\big((\alpha-1)+n(z+1)\big)^2-(\alpha-1)\}\bigg)\bigg]\nonumber\\
&/& 2 \bigg[(\Omega_{\phi 0}+\Omega _{m0}-1) (z+1)^{2 k}- \Omega_{\phi 0} e^{n z} (z+1)^{\alpha}-(z+1)^3\Omega _{m0}\bigg]
\end{eqnarray}
\begin{eqnarray}
s & = & \frac{r-1}{3 (q-\frac{1}{2})}\nonumber\\
 & = & \bigg[2k(3-2 k)(z+1)^{2 k}\left(\Omega_{m0}+\Omega_{\phi0}-1\right)+e^{n z} (z+1)^{\alpha} \Omega_{\phi 0} \bigg(\{n(z+1)+(\alpha-)\}^2-(\alpha+1)\bigg)\bigg]\nonumber\\
& / & 3\bigg[(3-2 k) (z+1)^{2 k} \left(\Omega _{m0}+\Omega _{\phi 0}-1\right)+e^{n z} (z+1)^{\alpha } \Omega_{\phi 0} \{\alpha-3 +n (z+1)\}\bigg]
\end{eqnarray}
Figure 7(a) depicts, the suggested model's evolutionary behavior in $r-s$ plane utilizing the derived expressions for $r$ and $s$. For the suggested model, the present values of ($r, s$) are calculated as (1.09673, -0.028773) by taking the combined sample of observational datasets into account. Thus, the given model starts with the matter-dominated era (decelerating scenario) cross $\Lambda$CDM $(r = 1, s = 0)$ enter quintessence era $(r < 1, s > 0)$ and finally approaches toward Chaplygin gas model $(r > 1, s < 0)$  in late time. Therefore, the suggested model behaves like the standard $\Lambda$CDM in the present scenario. \\
 
Using the best plausible values of parameters estimated from a combined sample of observational datasets, we have plotted the ($q, r $) trajectory in Figure 6(b) for the proposed model. In the figure horizontal red line ($r=1$) indicates the $\Lambda$CDM line which divides the evolutionary plane into two regions namely the chaplygin gas ($r > 1$) and quintessence DE ($r < 1$). The ($q, r $) trajectory for our proposed model starts from SCDM (0.5, 1) and shows a chaplygin gas-like behavior in late time. The model exhibits a flipping from a deceleration era to an acceleration scenario as depicted by the trajectory in ($q-r $) plane.
%%%%%%%%%%%%%%%%%%%%%%%%%%%%%%%%%%%%
\subsection{Jerk Parameter}
Another diagnostic that is utilized widely in astrophysical studies is the jerk parameter ($j$). The cosmic jolt (or jerk) is the basic idea behind the concept of the jerk parameter that creates a transition of the universe to an accelerating scenario from the decelerating era. A physical jerk is the pace at which acceleration changes in relation to time. It is derived by using the third-order term of Tyalor's expansion of the scale factor about $a_0$ in cosmology. This parameter offers us an additional edge in identifying kinematically degraded cosmic models \cite{ref74}.  It provides greater accuracy in understanding the expansion of cosmos in comparison to the Hubble parameter because of the involvement of the scale factor’s third-order derivative. It is possible to define $j$ in the suggested model as \cite{ref75,ref76,ref76a,ref77,ref78}.
%%%%%%%%%%%%%%%%%%%%%%%
\begin{equation}
%j=1-(1+z) \frac{H'(z)}{H(z)}+\frac{1}{2}(1+z)^2 \left[\frac{H''(z)}{H(z)}\right]^2
j(z) = (2q+1) q+(z+1) \frac{d q}{dz}
\end{equation}
From Eqs. (35) and (38), the expression of jerk diagnostic is developed as

\begin{eqnarray}
j(z) &=&\bigg[2 \left(2 k^2-3 k+1\right) (z+1)^{2 k} \left(\Omega _{m0}+\Omega_{\phi 0}-1\right)-2 (z+1)^3 \Omega_{m0}\nonumber\\
&-&e^{n z} (z+1)^{\alpha } \Omega_{\phi 0}\bigg(((\alpha -1)+n (z+1))^2-(\alpha -1)\bigg)\bigg]\nonumber\\
&/& 2\bigg[ \left((z+1)^{2 k} \left(\Omega_{m0}+\Omega_{\phi 0}-1\right)-(z+1)^3 \Omega _{m0}-e^{n z} (z+1)^{\alpha } \Omega_{\phi 0}\right)\bigg]
\end{eqnarray}

%%%%%%%%%%%%%%%%%%%%%%%%%%%%%%%%%%%%%%%%%%%%%%%%%%%%%%%%%%%%%%%%%%%%%%%%

\begin{figure}
\centering
\includegraphics[scale=0.60]{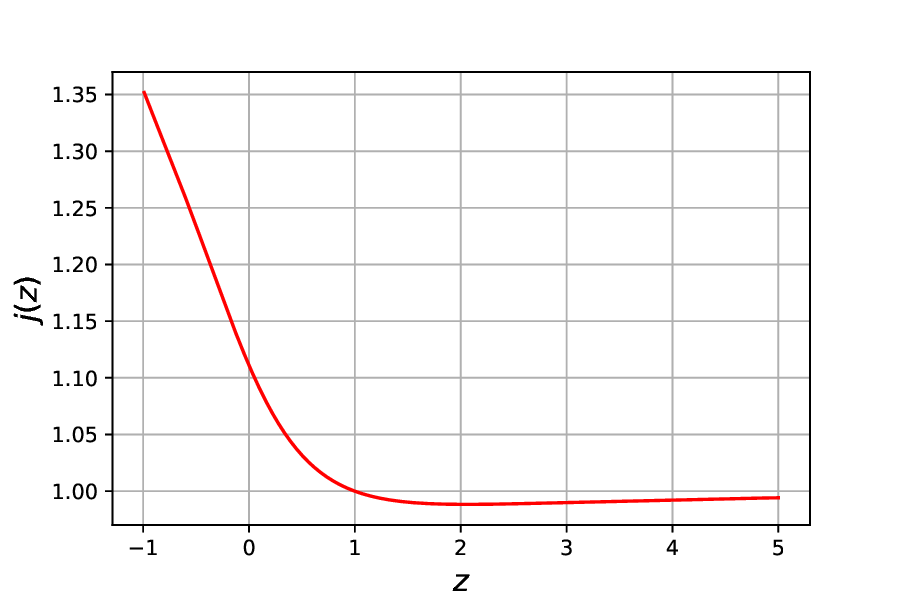}
\caption{Jerk paramter}
\end{figure}
%%%%%%%%%%%%%%%%%%%%%%%%%%%%%%%%%%%%%%%%%%%%
We have presented the graphical behavior of jerk parameter $j$ in Figure 8, using the best plausible values of free parameters estimated using the combined sample of Pantheon, BAO, and OHD datasets. The positive nature of the jerk parameter during the entire evolution indicates a smooth transition of the cosmos into an accelerated phase (see Fig. 8)\cite{ref76a,ref77,ref78}. The suggested model reflects the characteristics of the $\Lambda$CDM model of the universe because, in our study, the jerk parameter's current value is found to be $j_0 = 1.11^{+0.23}_{-0.16}$ which is $\approx 1$. It can be also observed from Figure, that our model acts like the standard $\Lambda$CDM model in the early time and represents another DE model that is distinct from $\Lambda$CDM in late time because $j_0 \neq 1$ \cite{ref75,ref76,ref76a,ref77,ref78}.
%%%%%%%%%%%%%%%%%%%%%%%%%%%%%%%%%%%%%%%% Table 2 %%%%%%%%%%%%%%%%%%%%%%%%%%%%%%%%%%%%%%%%%
\begin{table}
	\renewcommand\thetable{2}
	\caption{ The numerical findings of derived cosmological model for joint dataset }
	\begin{center}
		\begin{tabular}{|c|c|c|c|c|c|}
			\hline 
			\tiny Parameters & \tiny $q_{0}$ & \tiny $z_{t}$	& \tiny $\omega_{\phi 0}$ & \tiny $j_{0}$ & \tiny $\omega_{eff 0}$ \\
			\hline
			\tiny SN+BAO/CMB+OHD & \tiny $-0.625^{+0.067}_{-0.085}$ & \tiny $0.756^{+0.005}_{-0.015}$  & \tiny $-1.042^{+0.068}_{-0.069}$ & \tiny $1.11^{+0.043}_{-0.040}$ & \tiny $-0.750^{+0.045}_{-0.057}$\\
			\hline
		\end{tabular}
	\end{center}
\end{table}
%%%%%%%%%%%%%%%%%%%%%%%%%%%%%%%%%%%%%%%%%%%%%%%%%%%%%%%%%%%%%%%%%%%%%%%%%%%%%%%%%%%%%%%%%%%%
\section{Concluding remarks}
In the current study, we investigated a scalar field cosmological model with Lyra's geometry to explain the current cosmic expansion in a flat FRW universe. We assumed a variable displacement vector as a component of Lyra's geometry. In the context of the conventional theory of gravity, we suggest a suitable parameterization of the scalar fields dark energy density in the hybrid function of redshift $z$, confirming the essential transition behavior of the universe. The main highlights are summarized as follows:
\begin{itemize}

\item  We present constraints on model parameters using the most recent observational data sets from OHD, BAO/CMB, and Pantheon and taking Markov Chain Monte Carlo (MCMC) analysis into account. For the proposed model, the best estimated values of parameters for the combined dataset (OHD, BAO/CMB, and Pantheon) are $ H_0 = 71.15\pm 0.26$ km/s/Mpc, $ \Omega_{m0}=0.2625\pm 0.0024$, $ \Omega_{\phi0} = 0.676\pm0.038$, $ \alpha=-0.22\pm0.13$, $n = 0.096\pm0.079$, and $k = 0.38\pm0.32$.

\item The model exhibits a flipping nature, and the redshift transition occurs at $z_t = 0.756^{+0.005}_{-0.015}$ and the current value of the decelerated parameter for the proposed model is calculated as $q_0 = -0.625^{+0.067}_{-0.085}$ for the combined dataset.

\item In our investigation, we have validated the quintom behavior of the model by establishing the current values of the scalar field EoS parameter. The current value of the scalar field EoS parameter is estimated as $\omega_{\phi 0} = -1.042^{+0.068}_{-0.069}$ for combined datasets. From the trajectory of $\omega_{eff}$, it has been detected that the model stays in the quintessence era during the evolution of the cosmos and approaches to a phantom scenario in late time. Hence, the profile of $\omega_{eff}$ depicts a quintom-like behavior of the cosmos.

\item The total density parameter for the proposed model tends to unity in late time i.e., at the current era. Thus, in the present study, a scalar field is taken as a substitute for DE to describe the accelerated expansion of the cosmos.

\item The nature of kinetic energy is positive and decreasing in the evolution of the universe as depicted in the figure. Thus, the derived model shows a quintessence-like behavior in which massless scalar field $\phi$ can be assumed with a potential $V(\phi)$ that mimics the gravitational field and efficiently describes the current inflation of the universe \cite{ref24,ref25,ref26}.

\item The given model starts with the matter-dominated era (decelerating scenario) cross $\Lambda$CDM $(r = 1, s = 0)$ enter quintessence era $(r < 1, s > 0)$ and finally approaches toward Chaplygin gas model $(r > 1, s < 0)$  in late time. Therefore, the suggested model behaves like the standard $\Lambda$CDM in the present scenario.

\item The suggested model reflects the characteristics of the $\Lambda$CDM model of the universe because, in our study, the jerk parameter's current value is found to be $j_0 = 1.11^{+0.23}_{-0.16}$. It can be also observed from Figure 8 that our model acts like the standard $\Lambda$CDM model in the early time and represents another DE model that is distinct from $\Lambda$CDM in late time because $j_0 \neq 1$.

\end{itemize}
Hence, the current study describes a model of a transitioning universe using the scalar field as a substitute for dark energy. Recent findings support the outcomes of the proposed model. 

%%%%%%%%%%%%%%%%%%%%%%%%%%%%%%%%%%%%%
\section*{Acknowledgment}
\noindent The authors wish to place on record their sincere thanks to the reviewer(s) for illuminating suggestions that have significantly improved our manuscript in terms of research quality.
%%%%%%%%%%%%%%%%%%%%%%%%%%%%%%%%%%%%%%%%%%%%%%%%%%%%%%%%%%%%%%%%%%%%%%%%%%%%%%%%%%%%%%%%%%%%%%%%%%%%%%%%%%%%%%%%%%%%%%%%%%%%%%%% 

%\section*{Data availability} All data generated or analysed during this study are included in this article. 
%\section*{Declaration of competing interest}
%The authors declare that they have no known competing financial interests or personal relationships that could have appeared to influence the work reported in this paper.	

\end{document}